\documentclass[11pt]{article}

\usepackage[letterpaper,textwidth=494.50888pt,vmargin=0.95in]{geometry}
\usepackage[T1]{fontenc}
\usepackage[utf8]{inputenc}
\usepackage{textcomp}
\usepackage{amsmath,amsfonts,amssymb}
\usepackage{graphicx}
\graphicspath{{./}}
\usepackage{booktabs}
\usepackage[round,authoryear]{natbib}
\usepackage[font=small,labelfont=bf]{caption}
\usepackage[hidelinks]{hyperref}
\usepackage{url}

\title{BEEF: A Hybrid Integrated Photonic Simulation Framework}

\author{%
  Jacob M. Hiesener\thanks{\texttt{jhiesener4@gatech.edu}} \and
  Stephen E. Ralph\thanks{Corresponding author: \texttt{stephen.ralph@ece.gatech.edu}}\\[0.5em]
  \normalsize School of Electrical and Computer Engineering,\\
  \normalsize Georgia Institute of Technology, Atlanta, GA 30332, USA
}

\date{}

\begin{document}

\maketitle

\begin{abstract}
\noindent
A Python framework is presented that retains first-principles electromagnetic modeling while extending simulation to devices whose length and geometric complexity place them beyond the practical reach of either method alone. The \textbf{B}ridged \textbf{E}ngine for eigenmode expansion (\textbf{E}ME) and finite-difference time-domain (\textbf{F}DTD) (BEEF) partitions a device into sections which are simulated using either EME or FDTD independently developed Maxwell solvers depending on what best suits each section's geometry, and cascades the resulting responses. BEEF calibrates the EME propagation constants against FDTD-measured phase accumulation, providing a mode identity verification diagnostic across solver interfaces. For a 2$\times$2 multimode-interference splitter, the hybrid result agrees with full-device FDTD to within 0.59\% (bar) and 0.76\% (cross) across the C-band; a tapered variant agrees to within 0.87\% and 2.01\%, respectively. For a directional coupler, effective-index calibration reduces cut-position dependence 167-fold. A circuit-level cascade reproduces the measured Fabry--Perot resonance lineshape and free spectral range while exploring the coupler-length and arm-length design plane through post-processing of four cached simulations. Finally, we use BEEF to simulate a 167.3~\textmu m-long 4$\times$4 MMI-based 90-degree optical hybrid that is computationally impractical for full-device FDTD, with a reduced-length variant confirming close agreement in transmission and relative phase.
\end{abstract}


\vspace{1em}

\section{Introduction}\label{sec:intro}

Finite-difference time-domain (FDTD) and eigenmode expansion (EME) solvers are widely used in integrated photonics for device simulation and design. FDTD directly solves Maxwell's time-dependent curl equations by discretizing space and time onto a grid and time-stepping the electric and magnetic fields forward in discrete increments \citep{yee1966numerical,taflove2005computational}. EME instead decomposes the structure into longitudinally-uniform cross-sections, solves for each cross-section's propagating eigenmodes, and couples them via mode-matching at the interfaces between sections \citep{bienstman2001optical}. FDTD is well suited to compact, geometrically complex regions such as tapers, bends, and mode-converting junctions where the field varies fully in three dimensions. Conversely, EME is well suited to long, waveguide-like regions such as multimode interferometers and directional coupler straight sections where propagation along a single axis dominates. Both methods have distinct advantages and disadvantages, and the choice of which to use depends on the specific device being simulated.

FDTD is a versatile, general-purpose method that can model arbitrary geometries and material distributions, but is computationally intensive and can become slow or infeasible for large-scale devices. Several of its accuracy and practicality limitations are inherent to the method itself rather than artifacts of insufficient computational resources \citep{taflove2005computational}. Increasing the resolution or simulation size increases the simulation duration steeply, roughly as the fourth power of resolution in 3D, since each spatial dimension's grid density and the number of time steps both scale with it. Additionally, the time step is set by the Courant stability condition \citep[Ch.~4]{taflove2005computational} on the grid spacing, which in turn is set by the smallest feature that must be resolved in the simulation, so a single small feature forces a fine time step across the entire domain. Devices such as MMIs illustrate this directly: although most of the device is a single, large multimode waveguide that would tolerate a coarse time step on its own, the small input and output tapers at its edges still set the time step for the entire simulation.

Numerical dispersion and staircasing errors are inherent inaccuracies that accumulate over long propagation distances \citep{oskooi2010meep}, and a large number of time steps compounds this error accumulation. Numerical dispersion is a result of using finite-difference derivatives to model the continuous form of Maxwell's equations. Staircasing errors arise because a curved or diagonal material boundary is conventionally represented on the FDTD method's rectilinear grid by assigning each grid cell entirely to one material or the other, approximating the true smooth interface as a jagged staircase of orthogonal steps; subpixel smoothing, which instead gives each boundary cell an effective permittivity, substantially mitigates this and restores the method's nominal second-order convergence \citep{farjadpour2006improving}. Additionally, high-dynamic-range simulations, such as those involving weakly-coupled waveguides, are limited by the solver's numerical noise floor, and high-$Q$ resonant structures require long simulation times to resolve their resonant modes \citep{oskooi2010meep}. Inverse design techniques compound the cost further, since gradient-based optimization typically requires hundreds to thousands of FDTD simulations to converge to an optimized design \citep{molesky2018inverse,hiesener2025seeded,hiesener2026enhancing}. Confining those solves to the critical regions that genuinely require them is therefore as much a prerequisite for inverse design as it is for conventional simulations.

EME, by contrast, is a more efficient method for simulating larger waveguide-based devices, but is limited to structures that can be decomposed into a series of uniform waveguide sections \citep{bienstman2001optical}. It struggles with sharp diagonal or curved interfaces, such as those found in bends or the tips of inverse tapers, and is not well suited to devices with complex three-dimensional features such as grating couplers or mode-converting junctions. EME can also be less accurate than FDTD for devices whose cross-section varies continuously, since it must be approximated as a series of discrete, piecewise-uniform sections. These complementary limitations mean that neither EME nor FDTD alone is suitable for simulating every device of interest, motivating a hybrid approach that leverages the strengths of both methods.

The \textbf{B}ridged \textbf{E}ngine for \textbf{E}ME and \textbf{F}DTD (BEEF) is a hybrid-methodology orchestration layer on top of existing FDTD and EME solvers. BEEF builds on two open-source electromagnetic solvers: Meep, an FDTD simulation package \citep{oskooi2010meep}, and MEOW, an EME solver for waveguide modeling \citep{meow}. Neither backend is fixed: any solver implementing the same sectioning interface can occupy its slot in the cascade, including GPU-accelerated FDTD implementations. The framework partitions an integrated photonic device into multiple sections, each simulated with whichever solver best suits its geometry, and combines the per-section results into a complete simulation of the full device. This enables efficient simulation of devices that mix compact, geometrically complex regions poorly suited to EME with long, uniform regions too costly to simulate with FDTD alone -- exactly the devices neither method can handle on its own. BEEF is implemented in Python; device geometries, simulation parameters, and solver settings are all defined directly in the same scripting environment.

Prior work has explored dividing an integrated photonic device or circuit into sections handled differently and cascading the results: \citet{samolis2015hybrid} decompose a single FDTD simulation into smaller blocks joined by modal transfer matrices, trading one large time-domain run for several smaller ones rather than substituting a different method for any section, while \citet{mingaleev2011hybrid} split a full photonic integrated circuit into frequency-domain-modeled passive clusters and time-domain-modeled interconnects and active elements. Other work performs per-section substitution closer to BEEF's own: \citet{stoffer2004comparison} divide a microresonator into its ring and its two bent-straight couplers and model each region with an analytic coupled-mode theory, validated against FDTD rather than combined with it. \citet{oktay2024computationally} replace the per-design mode solve with lookups from a precomputed database of waveguide eigenmodes and overlaps, cascading the retrieved scattering matrices to obtain the device response and validating the result against 3D FDTD. BEEF instead cascades two independently developed, general-purpose electromagnetic solvers, neither an analytic surrogate nor a decomposition of the same solver, and provides a diagnostic that checks that the two engines' independently built interface modes actually agree with each other before contracting them together (Section~\ref{sec:mode_identity_verification}).

\begin{figure}[tbp]
\centering
\includegraphics[width=238.25444pt]{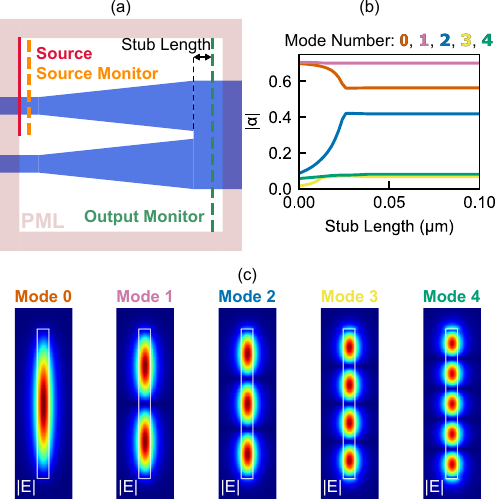}
\caption{Demonstrating the feasibility of a cascaded framework: light settles into a multimode region's modes within a few hundredths of a micron, so a section boundary placed just inside the body carries a well-defined modal basis. (a) FDTD simulation of the front end of a 2$\times$2 multimode interference (MMI) splitter, showing the source, source monitor, output monitor -- offset from the end of the taper by the stub length -- and perfectly matched layer (PML) placement. (b) Sweeping the stub length from 0 to 0.10 \textmu m in 0.002 \textmu m increments (40 pixels/\textmu m resolution), the modal overlap coefficient magnitude $|\alpha|$ measured at the output monitor converges within a few hundredths of a micron, confirming that a short stub is sufficient to reach a stable modal decomposition at the cut. Fourteen modes are carried at this interface; the five lowest-order guided modes are plotted. (c) Electric field magnitude $|E|$ of each of these five modes.}
\label{fig:stub_sweep}
\end{figure}

This sectioning approach is possible because many photonic devices have well-defined modes that dominate their behavior over large sections of their length \citep{snyder1983optical,bienstman2001optical}. For example, light entering the multimode region of a 2$\times$2 splitter settles into the modes of that region within the first few hundredths of a micron (Fig.~\ref{fig:stub_sweep}), demonstrated by sweeping the location of an output monitor and measuring the modal overlap coefficients of the first five modes in an FDTD simulation (Fig.~\ref{fig:stub_sweep}a,b). The FDTD monitors capture this phase and amplitude directly; the mode field profiles, if needed for further simulation, are instead computed with a separate mode solver (Fig.~\ref{fig:stub_sweep}c). Because these two engines solve for the same modes independently, an essential requirement of this framework is that the modes measured by the FDTD monitors are correctly correlated with those computed by the mode solver, addressed directly in Section~\ref{sec:mode_identity_verification}. Beyond MMI devices, this same sectioning approach extends to directional couplers, tapers, and any other device whose modes are well defined over a large portion of its length.

In this manuscript, we describe the BEEF simulation framework, including the device geometry interface (Section~\ref{sec:geometries}), the material models (Section~\ref{sec:material_models}), the solvers included in the BEEF cascade (Section~\ref{sec:solvers}), how mode identity is verified across solvers (Section~\ref{sec:mode_identity_verification}), and general implementation notes of the engine (Section~\ref{sec:implementation_notes}). We then describe four example applications: a standard 220~nm silicon-on-insulator (SOI) 2$\times$2 MMI splitter (Section~\ref{sec:2x2_mmi_splitter}), a directional coupler (Section~\ref{sec:dc}), a Fabry-Perot resonator built from it (Section~\ref{sec:hcg_fp}), and a 4$\times$4 MMI hybrid (Section~\ref{sec:4x4_mmi_hybrid}). Finally, we make concluding remarks in Section~\ref{sec:conclusions}.

\section{Simulation Paradigm}\label{sec:simulation_paradigm}

BEEF is a generic hybrid simulation framework that orchestrates the execution of independently-developed FDTD and EME solvers. The framework establishes a general cascade of these solvers through a shared contract (i.e. identical input and output modes, geometries, material models, simulation wavelengths, etc.) rather than through solver-specific code. The \texttt{solve\_section} method, implemented by every solver in the cascade, is the shared contract that allows any solver implementing it to occupy any position in the cascade, in any order, and any count. \texttt{HybridSolver} never branches on solver type: where a section's behaviour must differ, it is selected by asking whether that solver implements a given capability rather than by testing what class it is, so a new solver need only provide \texttt{solve\_section} to participate. We also include roughly a dozen further optional methods, unlocking features such as on-disk caching, propagation-constant calibration, and analytic length sweeps. This allows for a flexible and extensible simulation paradigm that can accommodate a wide range of photonic devices.

\subsection{Device Geometry Interface}\label{sec:geometries}

The \texttt{DeviceGeometry} class provides a common interface for defining the geometry of photonic devices. A device defines its shape once, as a set of polygons, from which both the prism representation the FDTD solver requires and the layer-based representation the EME solver requires are derived automatically; a device may alternatively be supplied directly as a GDSII layout. Both solvers are served from a single geometric source, rather than from two independently maintained descriptions, which is what makes the cross-solver consistency checks of Section~\ref{sec:mode_identity_verification} meaningful. The methods of this class allow any solver to query the geometry between the start and end of a specified section of the cascade. The \texttt{width\_at} and \texttt{center\_at} methods allow for the determination of the width and center of the device at any point along the section, enabling consistent and accurate simulation of the device geometry across both solvers.

One requirement of the engine is that the \texttt{section\_prisms} and \texttt{section\_gds} methods require the start and end of each section to land in a constant-cross-section region; cutting inside a varying region such as a taper raises a value error, as the mode in that section is not well-defined. Various device geometries are built into the framework, including multimode interference devices (MMIs), directional couplers, tapers, S-bends, uniform multi-waveguide cross-sections, and the bend and coupler sections of a ring resonator. A \texttt{CompoundGeometry} class concatenates any of these along the propagation axis, translating each section query into the frame of whichever component owns it, so that a device assembled from several parts is cascaded as a single chain; this is how devices requiring many more than three sections are constructed. Additionally, the framework allows for the definition of custom device geometries either by subclassing \texttt{DeviceGeometry} or through the use of GDSII files, enabling the simulation of a wide range of photonic devices.

\subsection{Material Models}\label{sec:material_models}

Identical Sellmeier models are used in both the FDTD and EME solvers to ensure the refractive index is consistent across both solvers. Currently three material models are implemented in the framework: silicon, using the Sellmeier fit of \citet{tatian1984fitting} to the measurements of \citep{salzberg1957infrared}, silicon dioxide via the Malitson model \citep{malitson1965interspecimen}, and silicon nitride via the Luke model \citep{luke2015broadband}. Additional material models can be added by extending the shared Sellmeier coefficient table, from which both solvers' native material representations are derived automatically.

Three pre-built stacks are implemented in the current framework: a 220 nm SOI stack, a 160 nm SOI stack, and a 300 nm silicon nitride stack. Because both solvers read the same analytic model, their refractive index \emph{lookups} at any given wavelength agree to machine precision. The media the two engines build from those lookups are not identical, however: the EME side evaluates the Sellmeier model exactly at each simulated wavelength, while the FDTD side must hand Meep a single dispersive medium valid across the whole simulated band at once, rather than a fresh index at every wavelength. Meep's medium is built by re-fitting the shared Sellmeier model to a single Lorentzian resonance over the simulated band, a fit that departs from exact Sellmeier by roughly $10^{-3}$ in index at the band edges. To catch a mismatched configuration before it contaminates a result, \texttt{HybridSolver} compares the index each solver actually uses at every section boundary and raises a warning if the two differ by more than $2\times10^{-3}$, twice the re-fit's own worst-case departure. Keeping this discrepancy an order of magnitude below the mode-solver differences discussed in Section~\ref{sec:solvers} is what allows the remaining disagreement between the two engines to be attributed to their discretization rather than to their materials. It is important to note that this tolerance measures the two engines' agreement with each other, not the absolute accuracy of the underlying material model, which is determined by the accuracy of the Sellmeier coefficients.

\subsection{Solvers: FDTD, EME, and the Hybrid Cascade}\label{sec:solvers}

\begin{figure}[tbp]
\centering
\includegraphics[width=238.25444pt]{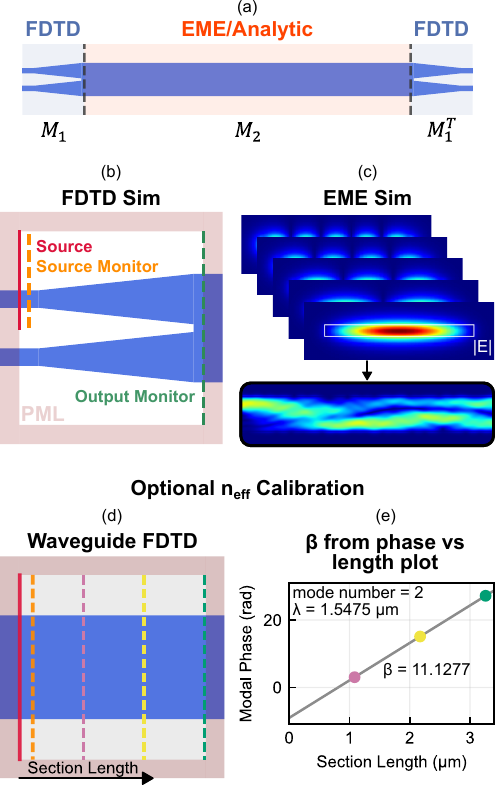}
\caption{Overview of the hybrid cascade implemented in BEEF, applied to a 2$\times$2 MMI splitter, outlining the partitioning of the geometry through the per-section solves and the optional propagation-constant calibration. (a) The device is partitioned into three sections: an input FDTD region, a multimode EME/analytic body, and an output FDTD region with transfer matrices $M_1$, $M_2$, and $M_1^T$, respectively. Because the device is mirror-symmetric, reciprocity gives the output section's transfer matrix as the transpose of the input section's rather than requiring it to be solved directly. (b) The FDTD simulation, showing the source, source monitor, output monitor, and PML placement. (c) The modes computed via the MIT Photonic Bands (MPB) mode solver and the transverse electric field magnitude $|E|$ computed by the MEOW EME simulation of the multimode body. (d)-(e) Optional $n_{\text{eff}}$ calibration: (d) short waveguide FDTD simulations of the body's own cross-section are run with monitors at different lengths, and (e) the resulting modal phase is fit against section length to extract a time-domain measurement of the propagation constant $\beta$ for each mode.}
\label{fig:hybrid_cascade}
\end{figure}

This framework relies on wrappers for the Meep and MEOW solvers, implemented using the \texttt{FDTDSolver} and \texttt{EMESolver} classes, respectively. Both classes can be applied to any section of the device geometry, provided the boundary between two sections falls at a constant, untapered cross-section (Fig.~\ref{fig:hybrid_cascade}a). The \texttt{FDTDSolver} class uses either a 2D planar FDTD simulation (for fast testing) or a full 3D FDTD simulation to solve the electromagnetic fields in a section's geometry (Fig.~\ref{fig:hybrid_cascade}b). The \texttt{EMESolver} class uses the eigenmode expansion method implemented in MEOW, with its mode solve routed either through MIT Photonic Bands (MPB) \citep{johnson2001blockiterative} via Meep \citep{oskooi2010meep} or through the Tidy3D mode solver \citep{tidy3d2022} (Fig.~\ref{fig:hybrid_cascade}c). Each solver sizes its own transverse mode window independently, through \texttt{sim\_width}/\texttt{sim\_height} on the FDTD side and \texttt{oxide\_width}/\texttt{oxide\_height} on the EME side; setting the two to match, as every example here does, is what keeps the interface bases directly comparable. These solvers can be used independently to simulate photonic devices, but the advantage of this framework is the ability to create a hybrid cascade.

By reciprocity, to determine the transfer matrix for a section, that section can be excited from whichever side carries fewer modes; \texttt{solve\_section} selects this side automatically, minimizing the number of simulations required. \texttt{HybridSolver} assembles the resulting section-wise transfer matrices into a single full-device response, generalizing the following three-section mode-overlap product, in which $\alpha^{(k)}_{m,n}$ is the modal overlap coefficient contributed by section $k$ -- an input FDTD region ($k=1$), an EME multimode body ($k=2$), and an output FDTD region ($k=3$, Fig.~\ref{fig:hybrid_cascade}a) -- for mode $n$ of the $N$ modes carried at the section interfaces, at frequency point $m$:

\begin{equation}
\alpha_m^{\text{full}} = \sum_{n=1}^{N} \alpha^{\text{(1)}}_{m,n} \, \alpha^{\text{(2)}}_{m,n} \, \alpha^{\text{(3)}}_{m,n}, \qquad T_m = \left| \alpha_m^{\text{full}} \right|^2.
\label{eq:scalar_cascade}
\end{equation}

to an arbitrary number of sections $K \geq 2$ with non-diagonal per-section transfer matrices $M_k[m] \in \mathbb{C}^{n_{\text{out},k} \times n_{\text{in},k}}$:

\begin{equation}
\text{full}[m] = M_K[m] \, M_{K-1}[m] \cdots M_1[m],
\label{eq:matrix_cascade}
\end{equation}

where the product is well-defined so long as $n_{\text{out},k} = n_{\text{in},k+1}$ at each interior interface -- an interior section, such as a multimode body, may therefore carry more modes than the sections on either side of it. Eq.~\eqref{eq:scalar_cascade} is the special case of a single-mode input and output bridged by an $N$-mode body: $M_1$ is the $N \times 1$ column coupling the input mode into the $N$ body modes, $M_2$ is the diagonal $N \times N$ matrix applying each body mode's propagation phase, and $M_3$ is the $1 \times N$ row coupling them back out, so their product is the $1 \times 1$ coherent sum of Eq.~\eqref{eq:scalar_cascade}. Eq.~\eqref{eq:matrix_cascade} relaxes both restrictions, admitting multiple modes at the device ports and off-diagonal coupling within a section.

For a device whose input and output sections are mirror images of one another -- a 2$\times$2 MMI splitter with mirror symmetry, for example -- \texttt{HybridSolver} can avoid resolving the mirrored section a second time by declaring \texttt{mirror\_pairs}, reusing section $j$'s transfer matrix, transposed over its mode axes, as section $k$'s result:

\begin{equation}
M_k[m] = M_j[m]^{T},
\label{eq:mirror}
\end{equation}

which, for the three-section case above, reduces the cascade to $\text{full}[m] = M_{\text{1}}[m]^{T} M_{\text{2}}[m] M_{\text{1}}[m]$ (Fig.~\ref{fig:hybrid_cascade}a). Unlike Meep's \texttt{mp.Mirror} symmetry \citep{oskooi2010meep}, a static property declared once when the simulation is constructed, \texttt{mirror\_pairs} is instead re-validated on every \texttt{solve()} call. This is necessary because \texttt{geometry\_kwargs} such as \texttt{input\_idx}/\texttt{output\_idx} can change which physical port is excited between calls, so a mirror relationship that holds for one call's port selection is not guaranteed to hold for another.

When MEOW's mode solve is routed through MPB, the eigenmode index MPB reports and the phase Meep's Yee-grid time-stepper actually accumulates disagree for the same physical cross-section. For example, at a grid resolution of 40 pixels/\textmu m a directional coupler's first even supermode has $\beta = 9.423$ rad/\textmu m from MPB and $\beta = 9.633$ rad/\textmu m from Meep's time-stepper. Both the even and the odd supermode are biased in the same direction, by $2.23\%$ and $2.36\%$ respectively, so the error in their difference is only $0.13$ percentage points -- but the coupling ratio depends on $\Delta\beta$ rather than on $\beta$, and with $\beta/\Delta\beta \approx 99$ for this device that small common bias is levered into a $10.4\%$ error in $\Delta\beta$. Removing it by mesh refinement alone would require a resolution of roughly 143 pixels/\textmu m, which for a 3D simulation is some two orders of magnitude more expensive than the resolution the device is otherwise converged at.

\texttt{HybridSolver} instead addresses this with the \texttt{calibrate\_\allowbreak body\_\allowbreak neff} flag: for each mode, it runs a short FDTD simulation of the EME body's own cross-section and fits the resulting phase-vs-length slope at the simulated wavelengths, measuring $\beta$ in the time domain for every mode at each wavelength (Fig.~\ref{fig:hybrid_cascade}d,e). The EME body's $n_{\text{eff}}$ is then overridden with this value, so both solvers agree on how fast phase accumulates through the body. The intent is self-consistency across the cut rather than an appeal to either engine's absolute accuracy; which of the two is closer to Maxwell's equations for a given cross-section is a separate question this framework does not settle. What can be verified is that the cascade stops depending on where the cut was made: for the directional coupler, sweeping the FDTD/EME boundary drifts the extracted $50{:}50$ coupling length by $0.109$ \textmu m per \textmu m of cut displacement uncalibrated, and by $-6.5\times10^{-4}$ \textmu m/\textmu m calibrated.

This disagreement is not unique to this framework. \citet{tsakmakidis2005systematic} extract the propagation constant of a strongly guiding rectangular waveguide from 3D FDTD by the same phase-against-distance fit, and report its error relative to a full-vectorial mode solver taken as the reference; they also find that the classical Yee scheme needs $6.5\times$ more grid points to reach the accuracy their nonstandard FDTD variant achieves. The sign we observe is the expected one: the Yee grid's numerical phase velocity decreases as the discretization coarsens \citep{schneider1999fdtd}, which raises the propagation constant the time-stepper accumulates relative to the mode solver's.

For a body represented by a single uniform section, the resulting transfer matrix is fully determined by these calibrated propagation constants alone, so \texttt{EMESolver} can skip the MPB mode solve; since that solve dominates the cost of propagating a section, this makes calibration substantially cheaper than the mode solve it replaces. A body spanning more than one internal cross-section (such as a tapered multimode region of a MMI) still requires the mode field profiles for EME propagation, and computes them as before.

A key advantage of this simulation framework is the ability to sweep the length of an EME/analytic section without additional FDTD or mode solves. Using the 2$\times$2 MMI splitter as an example, changing the length of the multimode section has no effect on the input and output FDTD sections and only affects the response of the multimode section itself, which is easily recomputed from the same mode solve at each swept length by re-applying the propagation phase. The \texttt{solve\_length\_sweep()} method implements this in the BEEF framework, simulating any range of lengths at the same computational cost as a single \texttt{solve()} call. Because the swept length enters only as an analytic phase, this is valid only over a constant cross-section; the framework samples the swept span and refuses the sweep if the width or center varies within it, rather than extrapolating a phase that has no meaning through a taper.

One limitation of this framework's cascade follows from Eq.~\eqref{eq:matrix_cascade} itself: the per-section matrices are one-directional, so light is transmitted forward from section to section, as though every interface were perfectly index-matched, and no reflected amplitude is returned to an upstream section. This is a property of BEEF's own cascade rather than of EME as a method: a fully bidirectional EME formulation can capture the reflection at each interface directly, which is what makes EME well suited to strongly-reflective periodic structures such as Bragg gratings \citep{gallagher2003eigenmode}, but propagating a full bidirectional scattering matrix through every section of the cascade would require additional FDTDs exciting each of the output modes to accurately calculate the output reflection coefficients of the FDTD sections. Reflections internal to a single section are captured by whichever solver resolves that section, but reflections between sections are not. This is the appropriate approximation for the devices the cascade targets, whose sections are joined at constant, adiabatic cross-sections where little is reflected by construction, and it is the leading candidate whenever a hybrid result differs from a full-device FDTD reference (Section~\ref{sec:2x2_mmi_splitter}). Devices whose behaviour depends on multiple internal reflections, such as a resonant cavity, are instead composed with the circuit layer described in Section~\ref{sec:implementation_notes}.

\subsection{Mode Identity Verification}\label{sec:mode_identity_verification}

Mode identity verification confirms that mode index $i$ of one solver's interface basis and mode index $i$ of the adjoining solver's basis correspond to the same physical mode, via the normalized transverse-field overlap between the two engines' sampled mode fields:

\begin{equation}
\eta_{ij} = \frac{\left| \int_A \mathbf{E}_{t,i}^{(a)} \cdot \mathbf{E}_{t,j}^{(b)*} \, dA \right|}{\sqrt{\left(\int_A |\mathbf{E}_{t,i}^{(a)}|^2 \, dA\right) \left(\int_A |\mathbf{E}_{t,j}^{(b)}|^2 \, dA\right)}}.
\label{eq:overlap}
\end{equation}

This overlap intentionally compares only the transverse $\mathbf{E}$-field rather than the reciprocity-derived $\mathbf{E}\times\mathbf{H}$ overlap, since Meep and MEOW report $\mathbf{E}$ and $\mathbf{H}$ on mutually inconsistent absolute scales for the same physical mode so crossing one engine's $\mathbf{E}$ with the other's $\mathbf{H}$ introduces a spurious per-engine scale factor unrelated to whether the modes actually match, and drives the diagonal of $\eta$ toward zero on every device. Correspondence between the two bases is decided by a one-to-one assignment that maximizes total overlap across the full $\eta$ matrix (the Hungarian algorithm on $-\eta$) rather than by matching each mode to its nearest neighbor in $n_{\text{eff}}$, which fails once modes crowd together. On a 16-mode basis for a 2.8~\textmu m-wide multimode region on 160 nm SOI, for instance, $n_{\text{eff}}$ matching sent three separate Meep bands to the same single MEOW band.

That the overlap matrix flags such a case is the point of computing it, but the diagnosis it enables matters more than the reassignment: the underlying cause was not a genuine near-degeneracy but a duplicated eigenvector, Meep's mode solver having returned the same field for two different requested bands, with an off-diagonal self-overlap indistinguishable from unity where every other pair of bands in the same matrix overlapped by less than $0.13$. A basis containing a repeated eigenvector is rank-deficient, and no reassignment of indices can repair it; the framework therefore treats a failed correspondence check as a condition to report rather than to silently permute, and the underlying case was resolved by enlarging the transverse domain until the interface returned a clean one-to-one identity.

To obtain the mode fields used on the FDTD side, Meep's \texttt{get\_eigenmode()} function is called at the desired cross-section, which internally performs an MPB eigenmode solve to recover the same modal field pattern Meep itself uses for eigenmode decomposition at that monitor. MEOW can also route its own mode solves through MPB via Meep (\texttt{mode\_solver="meep"}), but the two solves are not identical: each engine builds its own independent geometry and mesh for the cross-section. \texttt{FDTDSolver} draws it from \texttt{section\_prisms()} within its own simulation domain and resolution, while \texttt{EMESolver} draws it from \texttt{section\_gds()} onto a mesh sized by its own, separately-set resolution and oxide dimensions. Because these two constructions can differ, mode identity verification is a genuine cross-check of two independently-discretized representations of the same physical mode, catching real geometry- and resolution-driven errors. The check is opt-in rather than automatic, since it costs an additional field sample at every interior interface and needs a straight run-out on both sides long enough to query without intruding on either FDTD section's absorbing boundary.

\subsection{Implementation Notes}\label{sec:implementation_notes}

Because this hybrid framework re-solves many of the same sections across sweeps, ports, and repeated runs, redundant mode and FDTD solves are common. Within the BEEF framework, this issue is addressed at two levels. An optional flag is passed to \texttt{solve()} that dedupes sections that repeat across excitations or calls, so a section shared between two port excitations is not solved more than once. Across runs, each solver caches to disk: if a \texttt{save\_dir} is supplied, every \texttt{solve\_section()} call is fingerprinted into a configuration record holding the parameters of the simulation, the call arguments, and the exact section geometry. The configuration is hashed to a subdirectory storing that record, the simulation result, and a rendering of the simulation geometry for user readability. A later call fingerprinting to the same hash is served from disk, with the stored record re-checked by full equality so that a hash collision or a change to the record's schema cannot return the wrong result. Two refinements have been added that make the cache hit more often than exact matching alone would allow. Firstly, if fewer interface modes than a cached entry is requested, a slice (along the mode axes) of that larger result rather can be taken than resimulating. Secondly, a \texttt{cache\_only} flag turns a miss into an error rather than a silent simulation, so re-executing a notebook for documentation can never launch a multi-hour solve unnoticed.

The engine includes plotting functions for viewing device geometries, calibration results, and simulation outputs. \texttt{plot\_hybrid\_geometry()} draws a device's polygons with each cascaded section labeled by whether FDTD or EME resolves it. The $n_{\text{eff}}$ calibration described in Section~\ref{sec:solvers} has its own built-in plotting functions for both the calibration geometry and the fitted phase-vs-length lines (Fig.~\ref{fig:hybrid_cascade}d,e). Plotting support continues to grow alongside the rest of the engine.

Because each device's overall response can also be represented as an S-parameter block, independently-solved devices can be composed the same way real photonic circuits are: with the Redheffer star product, which correctly accounts for the multiple internal reflections between two connected ports rather than assuming one-pass transmission. We implement this as a simple networking layer, allowing separately simulated components to be connected into larger integrated photonic circuits. This is a distinct composition mechanism from \texttt{HybridSolver}'s cascade: \texttt{HybridSolver} composes sections of one device end-to-end, while \texttt{Circuit} composes independently-solved devices into an arbitrary network topology, including closed loops. For a resonant cavity whose response is available in closed form, the network layer reproduces the analytic reflection and transmission to better than one part in $10^{15}$, so any disagreement in a composed circuit is attributable to the constituent devices' own S-parameters rather than to the composition.

Two further utilities check the two free parameters the hybrid method introduces, and are device-agnostic. The choice of where to cut is checked by solving the same device at a range of cut positions and reporting the drift of the resulting response with cut placement: a correct decomposition is invariant to it, so any systematic slope is a defect, and it is this measurement that quantifies the improvement from $n_{\text{eff}}$ calibration reported in Section~\ref{sec:solvers}. The choice of how many modes to carry at each interface is checked by increasing that count until the response plateaus, which tests whether the interface basis is complete rather than merely self-consistent. Both are reported as diagnostics rather than enforced as gates, since the appropriate tolerance is device-dependent.

\section{Example Applications}\label{sec:test_cases}

A library of examples has been developed using this simulation framework; a handful, shown here, each demonstrates a different capability of the BEEF engine. A basic 220~nm SOI 2$\times$2 MMI splitter demonstrates the accuracy of the \texttt{solve\_length\_sweep()} method against full FDTD simulations. The directional coupler and Fabry-Perot resonator examples demonstrate the network/circuit capability, cascading independently-solved devices into a closed resonant loop via the Redheffer star product. These examples are also compared with experimental data to validate the simulation accuracy. The 4$\times$4 MMI-based 90-degree hybrid is an exceedingly long device infeasible to simulate with FDTD alone, demonstrating BEEF's ability to simulate devices not otherwise possible.

\subsection{2$\times$2 MMI Splitter}\label{sec:2x2_mmi_splitter}

\begin{figure}[tbp]
\centering
\includegraphics[width=238.25444pt]{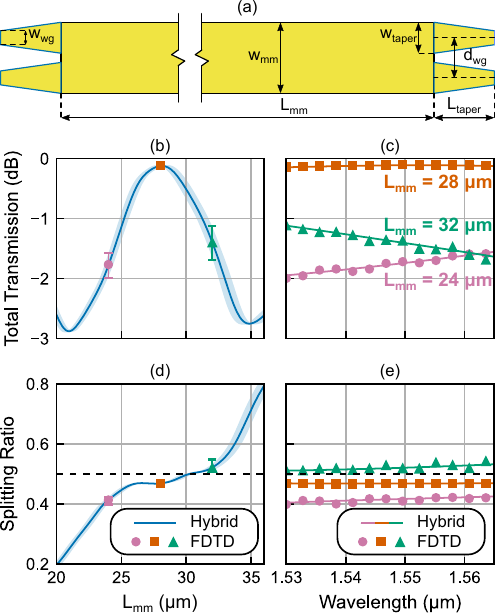}
\caption{Accuracy of the hybrid cascade applied to a 2$\times$2 MMI splitter validated against full-device FDTD. (a) The parameterized model of the 2$\times$2 splitter, with the multimode section length $L_{\text{mm}}$ marked; the remaining symbols are defined in Table~\ref{tab:mmi_2x2_params}; the four tapers (outlined in blue) share identical dimensions. (b) Total transmission, averaged over the C-band (1.530 to 1.565~\textmu m), as a function of $L_{\text{mm}}$, computed with \texttt{solve\_length\_sweep()} and validated against full 3D FDTD simulations at three lengths (markers). The solid line is the band average and the shaded region spans its minimum to maximum. (c) Total transmission vs. wavelength at each of the three validated lengths, comparing the hybrid result (lines) against full FDTD (markers). (d) Splitting ratio as a function of $L_{\text{mm}}$, same hybrid sweep and FDTD validation points as (b). (e) Splitting ratio vs. wavelength at the same three lengths, hybrid vs. full FDTD.}
\label{fig:mmi_2x2}
\end{figure}

MMI-based 2$\times$2 splitters are fundamental building blocks of many integrated photonic circuits, offering broadband, high transmission with an equal splitting ratio \citep{soldano1995optical}, and are themselves a frequent target of geometry optimization \citep{hiesener2026toward}. The splitter shown here targets a standard 220~nm SOI platform with a minimum feature size of 200~nm, making it feasible for almost any UV-lithography-based fabrication process. The device geometry is generated in the BEEF framework using the \texttt{MMI} device geometry class, and is defined by six parameters (Fig.~\ref{fig:mmi_2x2}a): waveguide width $w_{\text{wg}}$, multimode section width $w_{\text{mm}}$, multimode section length $L_{\text{mm}}$, taper width $w_{\text{taper}}$, taper length $L_{\text{taper}}$, and waveguide separation $d_{\text{wg}}$. Preliminary EME parameter sweeps were performed to arrive at the working values listed in Table~\ref{tab:mmi_2x2_params}.

\begin{table}[tbp]
\centering
\caption{Geometric parameters of the 2$\times$2 MMI splitter (Fig.~\ref{fig:mmi_2x2}a).}
\label{tab:mmi_2x2_params}
\begin{tabular}{lll}
\toprule
Symbol & Description & Value \\
\midrule
$w_{\text{wg}}$ & Waveguide width & 0.45~\textmu m \\
$w_{\text{mm}}$ & Multimode section width & 2.8~\textmu m \\
$L_{\text{mm}}$ & Multimode section length (nominal) & 28.0~\textmu m \\
$w_{\text{taper}}$ & Taper width & 1.3~\textmu m \\
$L_{\text{taper}}$ & Taper length & 4.0~\textmu m \\
$d_{\text{wg}}$ & Waveguide separation & 1.5~\textmu m \\
\bottomrule
\end{tabular}
\end{table}

The hybrid simulation for this example uses the \texttt{calibrate\_\allowbreak body\_\allowbreak neff} flag to overwrite the effective index of the multimode section's modes to match the FDTD-measured value. Unlike the directional coupler of Section~\ref{sec:dc}, this device does not require it: the two engines agree on the body's differential propagation constant to $0.47\%$ here, against roughly $10\%$ for the coupler, and the agreement varies monotonically across the C-band. The mechanism that makes calibration indispensable for a directional coupler is therefore essentially absent from this device, whose splitting depends on the modal beat length of a single wide multimode region rather than on the difference between two nearly-degenerate supermodes. Calibration is applied regardless, since it costs a short waveguide simulation and removes the question, but it is not what sets the accuracy of this device.

As described in Section~\ref{sec:solvers}, the device's mirror symmetry is exploited twice to reduce the number of FDTD simulations required. First, the output section is never separately simulated: by reciprocity, exciting output port $k$ is equivalent to exciting input port $k$ on the mirrored device, so the output section's transfer matrix is simply the transpose of the input section's. Second, launching the bottom input arm is itself a mirror operation of launching the top arm, so each body mode's coefficient either matches or flips sign between the two launches according to its even/odd parity across the mirror axis; this per-mode sign vector was measured once against a real two-arm FDTD run and reused thereafter, dropping the number of input-side FDTD simulations from two to one. Together, these reductions mean the entire hybrid response is obtained for any multimode section length from a single FDTD simulation, exciting one input waveguide and measuring all interface modes.

Three complete-device FDTD simulations were performed with multimode section lengths of 24, 28, and 32~\textmu m. Both the FDTD and hybrid simulations were performed using a grid resolution of 40 pixels/\textmu m. The FDTD transmission and splitting ratio are compared with the hybrid sweep in Fig.~\ref{fig:mmi_2x2}b,d, and the wavelength-resolved spectra at each of these three lengths are compared in Fig.~\ref{fig:mmi_2x2}c,e. The hybrid and full FDTD results agree closely across the C-band, with a maximum transmission discrepancy of 0.59\% (bar) and 0.76\% (cross). This discrepancy changes sign across the simulated band rather than acting as a uniform offset, which rules out an error in the body's propagation constant: a monotonic bias in $\Delta\beta$, such as the one calibration corrects, cannot produce a residual that crosses zero. It is instead consistent with the reflections at the taper transitions that the one-directional cascade of Section~\ref{sec:solvers} does not carry between sections. These discrepancies are nonetheless minuscule, and the hybrid simulation agrees closely with the full FDTD result at all three validated lengths.

\begin{figure}[tbp]
\centering
\includegraphics[width=238.25444pt]{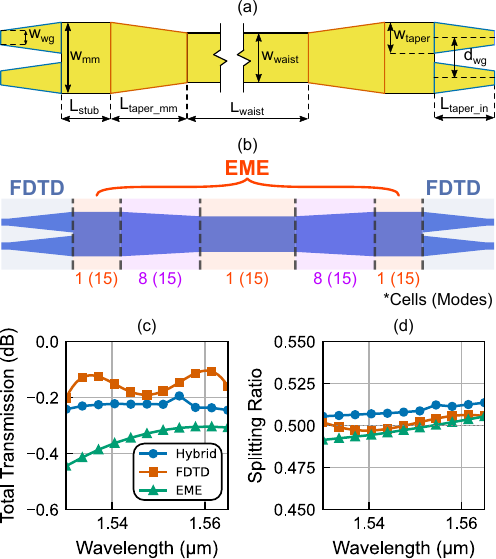}
\caption{Demonstration of the hybrid cascade applied to a 2$\times$2 MMI whose EME cross-section varies along its length. (a) The parameterized model of the tapered 2$\times$2 splitter, with the marked symbols defined in Table~\ref{tab:tapered_mmi_2x2_params}. The four input/output tapers (outlined in blue) share dimensions, as do the two multimode-section tapers connecting the waist to the outer multimode section (outlined in orange). (b) The hybrid solver's decomposition of the device: FDTD sections cover the input and output tapers, while the EME body is itself split into five sub-sections (stub, taper, waist, taper, stub) each solved with the number of cells and, in parentheses, the number of modes retained per cell labeled below. (c) Total transmission and (d) splitting ratio spectra of the tapered splitter when simulated with the hybrid method (blue circles), full-device FDTD (orange squares), and standalone EME (green triangles).}
\label{fig:tapered_mmi_2x2}
\end{figure}

A common design strategy for shortening a 2$\times$2 MMI splitter is to taper the multimode section down to a narrower waist along its length (Fig.~\ref{fig:tapered_mmi_2x2}a): since the beat length of the fundamental mode pair scales with the multimode width, narrowing the body shortens the beat length and, with it, the overall device length \citep{wan2007length}. Because the multimode body's cross-section varies continuously through the transition regions, it cannot be represented by a single uniform EME slab and must instead be divided into several distinct cells (Fig.~\ref{fig:tapered_mmi_2x2}b), making this device a natural demonstration of the hybrid solver's multi-cell cascade. As with the untapered 2$\times$2 splitter, the device is split into three sections (input taper, multimode body, and output taper) and the same output-section mirror symmetry applies: a single FDTD simulation captures the performance of both the input and output tapers. The multimode body is itself mirror-symmetric about its center line (Fig.~\ref{fig:tapered_mmi_2x2}a,b), and this symmetry is exploited a second time within the EME cascade, since the transition and stub cells nearest the output are identical in cross-section to their counterparts nearest the input. Only one cell of each mirrored pair requires an independent mode solve, halving the number of cross-sections that must be solved within the body.

The transmission and splitting ratio spectra of the tapered 2$\times$2 MMI splitter are simulated using the hybrid method, a full-device FDTD simulation, and a full-device EME simulation (Fig.~\ref{fig:tapered_mmi_2x2}c,d). The hybrid and full FDTD results agree closely, with a maximum transmission discrepancy of $0.87\%$ (bar) and $2.01\%$ (cross) -- larger than the untapered device's own $0.59\%$/$0.76\%$, consistent with the additional taper-transition reflections that this longer chain of sections introduces. The full-EME simulation, by contrast, departs sharply from both FDTD-anchored results, with a maximum transmission discrepancy of $3.69\%$ (bar) and $2.49\%$ (cross). The standalone EME must also represent the input and output tapers within a truncated guided-mode basis, and at the taper tip's abrupt junction with the multimode body a significant fraction of the field couples to radiation that the retained guided modes cannot express. The hybrid method avoids this by handing exactly those two regions to FDTD, which resolves the radiated and evanescent fields directly. The hybrid method therefore reproduces the full-device FDTD result far more faithfully than EME alone. The EME body's own transition regions are nonetheless still approximated as a staircase of uniform cells (Fig.~\ref{fig:tapered_mmi_2x2}b); a first-order taper-integration subcell method, which propagates the local mode coefficients along a smoothly varying taper rather than through a sequence of discrete cell interfaces, could reduce this residual staircase error without adding cells or FDTD simulations \citep{gallagher2003eigenmode}. Together with the untapered variant, this device demonstrates that the hybrid cascade retains its accuracy even as the multimode body's cross-section is allowed to vary along its length.

\begin{table}[tbp]
\centering
\caption{Geometric parameters of the tapered 2$\times$2 MMI splitter (Fig.~\ref{fig:tapered_mmi_2x2}a).}
\label{tab:tapered_mmi_2x2_params}
\begin{tabular}{lll}
\toprule
Symbol & Description & Value \\
\midrule
$w_{\text{wg}}$ & Waveguide width & 0.45~\textmu m \\
$w_{\text{mm}}$ & Multimode section width & 2.8~\textmu m \\
$w_{\text{waist}}$ & Waist width & 2.24~\textmu m \\
$w_{\text{taper}}$ & Input/output taper width & 1.3~\textmu m \\
$d_{\text{wg}}$ & Waveguide separation & 1.5~\textmu m \\
$L_{\text{stub}}$ & Uniform multimode stub length & 3.0~\textmu m \\
$L_{\text{taper,mm}}$ & Multimode-to-waist taper length & 5.0~\textmu m \\
$L_{\text{waist}}$ & Waist length & 6.0~\textmu m \\
$L_{\text{taper,in}}$ & Input/output taper length & 4.0~\textmu m \\
\bottomrule
\end{tabular}
\end{table}

\subsection{Directional Coupler}\label{sec:dc}

\begin{figure}[tbp]
\centering
\includegraphics[width=494.50888pt]{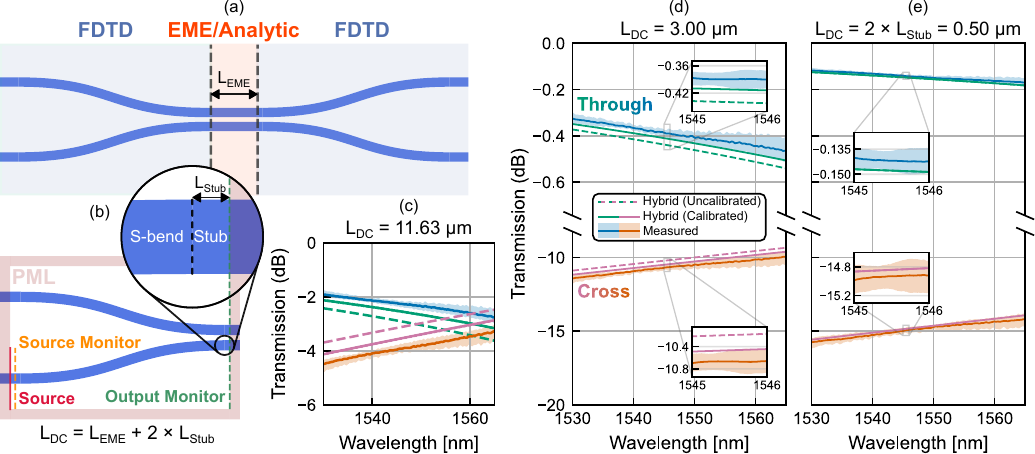}
\caption{The hybrid cascade applied to a directional coupler, compared against measurements at three coupling lengths and demonstrating the effect of effective-index calibration. (a) The hybrid cascade geometry for the directional coupler, partitioned like the 2$\times$2 MMI splitter, with the EME body length $L_{\text{EME}}$ marked. (b) The FDTD simulation, with a zoomed-in inset showing the stub length $L_{\text{stub}}$ between the end of the S-bend and the output monitor; the coupling length is $L_{\text{DC}} = L_{\text{EME}} + 2L_{\text{stub}}$. (c) $L_{\text{DC}} = 11.63$~\textmu m, the $50{:}50$ coupler; (d) $L_{\text{DC}} = 3.00$~\textmu m; and (e) $L_{\text{DC}} = 0.50$~\textmu m. Each panel shows through (blue: measured; green: simulated) and cross (orange: measured; pink: simulated) transmission, with the calibrated hybrid drawn solid and the uncalibrated hybrid dashed. For the measured data, the dark lines depict the mean values; the light band depicts the worst- and best-case performance across all measured chips.}
\label{fig:dc}
\end{figure}

Directional couplers are fundamental integrated photonic devices, using the evanescent coupling between closely-spaced waveguides to split light with negligible loss. The performance of these devices is dominated by the supermodes of the coupled waveguide pair, which are readily modeled with an EME or mode solver alone. Coupling into and out of the paired waveguides, however, requires bends or S-bends at the input and output, and these bending regions introduce coupling that is not captured by a mode solve over the straight coupled section. Directional couplers are typically designed by obtaining the coupling coefficient from mode simulations over the straight section and accounting for the bends either by adding an effective length extracted from measurement or by performing a full-device 3D FDTD simulation \citep{chrostowski2015silicon}. Instead, our hybrid methodology simulates the full device by resolving the input and output bend regions with FDTD and the straight coupled-waveguide length with EME, capturing both effects within a single cascade (Fig.~\ref{fig:dc}a). This device is also the first of two compared against measurement, validating the cascade against experiment along with full-device FDTD.

As with the MMIs, a stub length of 0.25 \textmu m ensures the FDTD-measured modes reach a stable state before the output monitor (Fig.~\ref{fig:dc}b). This device follows the same solver architecture as the 2$\times$2 MMI splitter: the coupled waveguide pair's own mirror symmetry means only its even and odd supermodes need to be carried through the EME body, and the device's longitudinal mirror symmetry lets the output section reuse the input section's transfer matrix, transposed, rather than being simulated directly. \texttt{HybridSolver} can express this parity relationship natively, declaring the cross path as a \texttt{MirrorSpec} carrying a measured mode-sign vector, which reduces a directional coupler to a single FDTD simulation, and the results reported below were generated this way, with the opposite excitation recovered from the first to a relative error below $10^{-4}$.

The directional couplers demonstrated here are fabricated on the 160~nm SOI GlobalFoundries (GF) Fotonix\textsuperscript{TM} 45~nm silicon photonics platform and use a gap size of 250~nm. Test structures for each coupler length were measured across multiple chips from two wafer samples to determine through and cross transmission. A standard fiber-array setup probed the GF process design kit (PDK) grating couplers used to access the directional coupler inputs and outputs, with loopback grating couplers connected by a single waveguide used to de-embed the grating coupler's own coupling loss. Wavelength sweeps were performed with a Keysight 8164B tunable laser and a Koheron PD10R photodiode, automated through the open-source photonic test framework LabExT~\citep{eppenberger2021labext}.

To test the hybrid solver's accuracy at arbitrary coupling lengths, we performed a hybrid length sweep: the FDTD S-bend excitation, its mirrored counterpart, and the FDTD-calibrated EME body mode solve are each computed once, and the body's per-length phase matrix is then swept analytically to give the through and cross transmission at any coupling length. Both the FDTD and EME body sections ran at 80 pixels/\textmu m. The resulting hybrid transmission is compared with the measured results, both calibrated and uncalibrated via the \texttt{calibrate\_\allowbreak body\_\allowbreak neff} flag on \texttt{HybridSolver}, at the three fabricated lengths (Fig.~\ref{fig:dc}c,d,e). Full single-domain FDTD simulations were also performed at the same lengths and the same resolution; since their maximum percent error against the calibrated hybrid simulation is only 0.15\% (Table~\ref{tab:dc_calibration}), the FDTD transmissions are omitted from Fig.~\ref{fig:dc} rather than plotted on top of an indistinguishable hybrid curve. This error is defined, for each transmission channel $X \in \{\text{through}, \text{cross}\}$, as the hybrid simulation's percent deviation from the full FDTD reference,

\begin{equation}
\varepsilon_X = \frac{T_X^{\text{hybrid}} - T_X^{\text{FDTD}}}{T_X^{\text{FDTD}}} \times 100\%,
\label{eq:dc_error}
\end{equation}

evaluated at each simulated wavelength and reported below as its worst case across the C-band.

\begin{table}[tbp]
\centering
\caption{Hybrid transmission deviation from full 3D FDTD for three fabricated directional coupler lengths, with and without $n_{\text{eff}}$ calibration (Section~\ref{sec:solvers}). Values are $\max_\lambda |\varepsilon_X|$ (Eq.~\eqref{eq:dc_error}) across the C-band.}
\label{tab:dc_calibration}
\begin{tabular}{lrrrr}
\toprule
& \multicolumn{2}{c}{Through} & \multicolumn{2}{c}{Cross} \\
$L_{\text{DC}}$ (\textmu m) & Uncal. & Cal. & Uncal. & Cal. \\
\midrule
0.50 & 0.02\% & 0.02\% & 0.15\% & 0.15\% \\
3.00 & 0.74\% & 0.02\% & 6.61\% & 0.15\% \\
11.63 & 9.60\% & 0.08\% & 10.55\% & 0.11\% \\
\bottomrule
\end{tabular}
\end{table}

The $n_{\text{eff}}$ calibration works as intended: it matches the hybrid result to the full single-domain FDTD result. For this device, calibration also improves alignment with the measured result, and that alignment grows closer as the coupling length is reduced (Fig.~\ref{fig:dc}c,d). At the shortest length, the coupling length equals twice the stub length, meaning the EME body length is zero (Fig.~\ref{fig:dc}e). The device therefore reduces to a cascade of two FDTD simulations, removing any need for $n_{\text{eff}}$ calibration. Though the simulations approach the measured result, a gap remains between the FDTD/calibrated hybrid simulations and the measurements, examined alongside the resonator's own discrepancy in Section~\ref{sec:hcg_fp}.

\subsection{Fabry-Perot Resonator}\label{sec:hcg_fp}

\begin{figure}[tbp]
\centering
\includegraphics[width=494.50888pt]{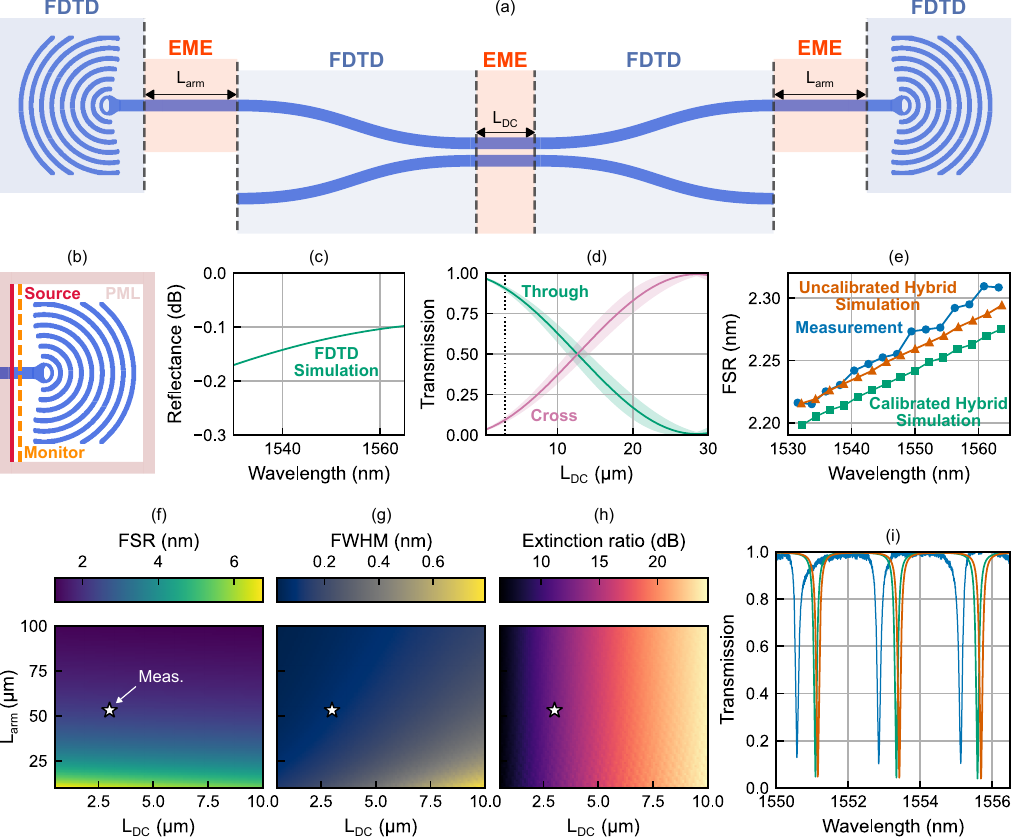}
\caption{Demonstration of the \texttt{circuit} method of BEEF on a Fabry-Perot resonator, reproducing the measured response. Additionally, the full design space of the resonator can be explored from a handful of cached simulations. (a) The hybrid cascade for the Fabry-Perot resonator: two high-contrast grating (HCG) mirrors (FDTD), two connecting-waveguide arms, and a directional coupler (FDTD S-bends and an EME body) joined into a closed cavity via the Redheffer star product, with the arm length $L_{\text{arm}}$ and coupling length $L_{\text{DC}}$ marked. (b) The HCG mirror's FDTD domain, built directly from its GDSII layout, with the source, monitor, and PML marked. (c) The resulting FDTD reflectance spectrum. (d) Through and cross transmission vs. $L_{\text{DC}}$ for the embedded coupler, from the same S-bend as Section~\ref{sec:dc}, reused analytically across coupling length via its own per-length phase matrix; the line is the C-band average and the shaded region spans its min to max, with the dotted line marking the fabricated device's own coupling length. (e) The measured free spectral range (FSR, blue) compared with the hybrid simulation with (green) and without (orange) $n_{\text{eff}}$ calibration of the connecting-waveguide arm. (f-h) FSR, full width at half maximum (FWHM), and extinction ratio across the ($L_{\text{DC}}$, $L_{\text{arm}}$) design plane, obtained as closed-form post-processing of four cached simulations: the HCG mirror, the coupler's S-bend excitation, its body beta calibrated mode simulation, and the arm's mode simulation with no additional mode or FDTD solves required. The white star in each heatmap marks the dimensions of the fabricated device. (i) The measured transmission spectrum (blue) near resonance compared with the hybrid simulations (green/orange).}
\label{fig:fp}
\end{figure}

The directional coupler is a fundamental building block of many integrated photonic systems. One such system is a Fabry-Perot resonator, which forms a resonant cavity used for filtering applications \citep{saber2018transversely}. One design method for this resonator uses high-contrast grating (HCG) reflectors on either end of the cavity, with a directional coupler tapping light in and out (Fig.~\ref{fig:fp}a). This HCG mirror was itself characterized as a test structure to measure its reflectance \citep{hiesener2025seeded,hiesener2024topology}. Using the hybrid cascade framework for the directional coupler from Section~\ref{sec:dc} and the \texttt{Circuit} class to connect each component, we simulate the entire Fabry-Perot resonator using 2 FDTD simulations (the HCG mirror and the directional coupler input) and 3 mode solves (the arm waveguide's fundamental mode and the directional coupler's two supermodes).

The HCG reflector is simulated by loading its geometry from a GDSII file and using BEEF's \texttt{FDTDSolver} class to perform an FDTD simulation at a resolution of 80 pixels/\textmu m (Fig.~\ref{fig:fp}b,c). The directional coupler is simulated using the hybrid method with a calibrated effective index, as described in Section~\ref{sec:dc}, to compute the through and cross transmission as a function of coupling length (Fig.~\ref{fig:fp}d). The arm waveguide's mode was computed with and without $n_{\text{eff}}$ calibration, and the \texttt{Circuit} class was used to combine all sections via the Redheffer star product. To avoid the thousands of mode simulations a directly-resolved spectrum would require, each section was instead simulated at 1~nm intervals and resampled onto a dense 20{,}001-point grid -- complex (magnitude/phase) interpolation for the HCG reflectance and the arm-to-body-mode overlap factors, and a degree-2 polynomial fit for the smoothly-varying body and arm propagation constants -- to reconstruct the resonance lineshape. 

The effect of calibrating the arm waveguide's effective index can be seen in Fig.~\ref{fig:fp}e, where the free-spectral range (FSR) of the calibrated and uncalibrated simulations is compared against the measurement; this comparison varies the arm alone, holding the coupler body calibrated throughout, so it isolates the connecting waveguide's own contribution to the cavity round trip. Since FSR is inversely proportional to effective index, the index shift between the two simulation variants translates directly into a shift in FSR. 

BEEF's \texttt{solve\_section\_length\_sweep()} method makes sweeping the directional coupler's coupling length essentially free: the coupled-waveguide body's supermode phases are evaluated analytically at any length from a single cached mode solve, without a new EME or FDTD run. The connecting arm's length enters the cavity response the same way, through its own closed-form propagation phase, so it too can be swept at no additional simulation cost. Together, these let the entire (coupling length, arm length) design plane be explored from the same handful of cached simulations: at each point we generate the transmission spectrum and reduce it to the resonator's headline performance metrics -- FSR, FWHM, and extinction ratio -- which we plot as heatmaps to quickly survey the design space (Fig.~\ref{fig:fp}f,g,h). 

The resonance peaks in the simulated resonator's transmission spectrum are offset from the measured result (Fig.~\ref{fig:fp}i); a $\sim$0.1\% reduction in the waveguide's effective index is enough to align them, consistent with the same fabrication-induced index deviation discussed above. The Lorentzian shape of the simulated resonance peaks matches the measured lineshape closely, with the small remaining deviation attributable to measurement noise and grating-coupler misalignment rather than to the simulation method itself.

Neither the Fabry-Perot resonator nor the directional coupler is reproduced exactly by any of the simulation variants. For the $50{:}50$ coupler at 1550~nm, $n_{\text{eff}}$ calibration halves the cross-transmission deviation from the measured value, from $0.82$ to $0.41$~dB, but it does so only by bringing the hybrid simulation to within $0.004$~dB of the full-device FDTD reference. The resonator exhibits the same mechanism with the opposite sign: its measured FSR of 2.261~nm is reproduced to 0.007~nm without $n_{\text{eff}}$ calibration and to 0.024~nm once the arm's index is replaced by the FDTD-measured one. This behavior is expected, as the $n_{\text{eff}}$ calibration enforces agreement between the two engines rather than agreement with the fabricated device. The two comparisons also exercise different parts of the cascade: Fig.~\ref{fig:dc} varies the coupled-waveguide body's propagation constants, while Fig.~\ref{fig:fp}e,i holds that body calibrated and varies only the connecting arm, with the HCG mirrors resolved by FDTD throughout and unaffected by either choice. Both residuals instead point to a difference between the simulated and measured geometry -- layer height, a non-vertical sidewall profile, or other fabrication realities that shift the waveguide's effective index and, with it, device performance \citep{lu2017performance}. Closing this gap requires extracting these parameters from dedicated test structures and correlating the as-built geometry against the simulated design, not a different solver.

This example is similar to a compact-model circuit simulation \citep{bogaerts2018silicon}; however, rather than solving Maxwell's equations for each component, a compact model represents every building block with a parameterized S-parameter or behavioral model, and cascades these analytic blocks through a circuit-level network solver to efficiently obtain the full system response. Our method instead derives each block's own S-parameters from full electromagnetic solves: FDTD for the HCG mirror and the coupler's bend regions, an EME mode solve calibrated against the FDTD-measured $\beta$ for the coupled-waveguide body and connecting arms. These solves are then cascaded with the same \texttt{Circuit} network machinery internal to the simulation framework. The result keeps a compact model's circuit-level scalability and low simulation cost while replacing its fitted or reduced-order component models with first-principles electromagnetic solutions of the real device geometry.

\subsection{4$\times$4 MMI 90-Degree Hybrid}\label{sec:4x4_mmi_hybrid}

\begin{figure}[tbp]
\centering
\includegraphics[width=494.50888pt]{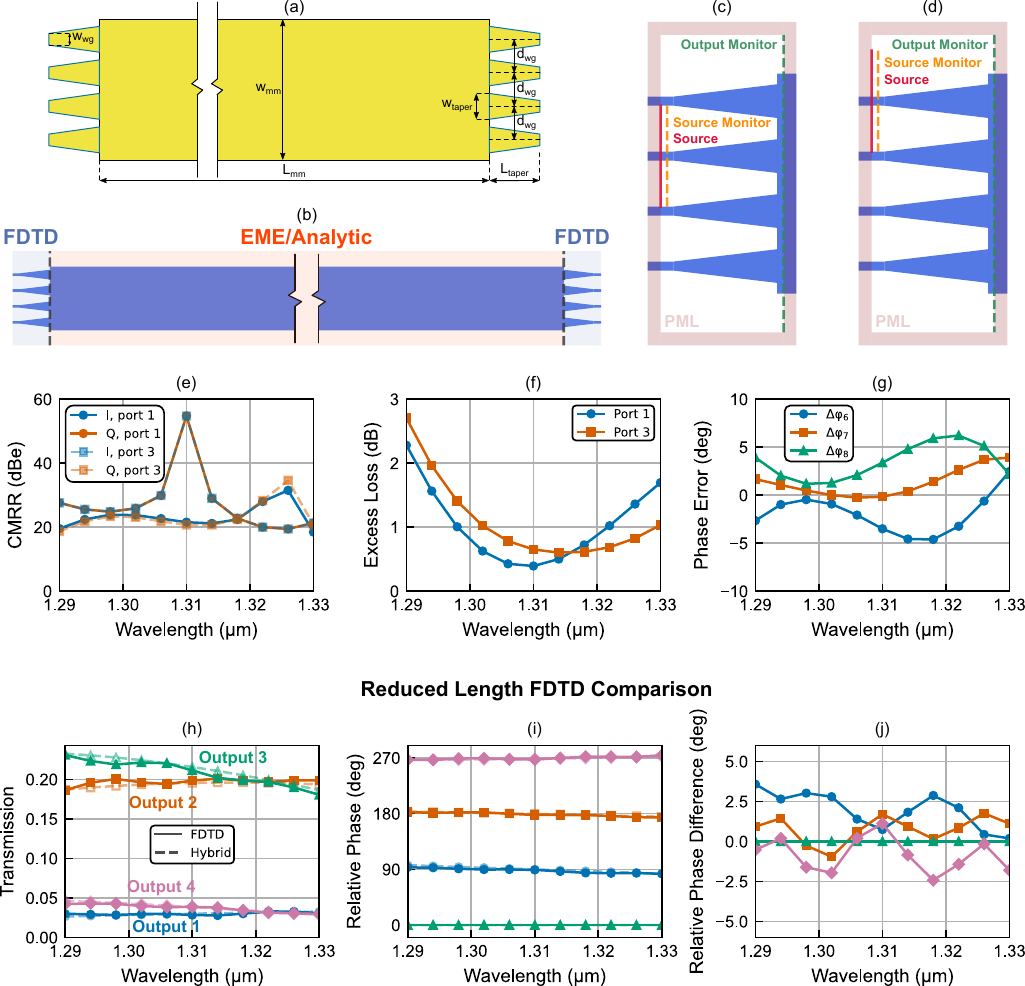}
\caption{The hybrid method applied to a device beyond the practical reach of full-device FDTD, with a reduced-length variant confirming agreement where a direct comparison remains affordable. (a) Parameterized geometry for the 4$\times$4 MMI 90-Degree Hybrid, with the multimode section length $L_{\text{mm}}$ marked; the remaining symbols are defined in Table~\ref{tab:mmi_4x4_params}. (b) The hybrid cascade: the multimode body is simulated with EME, and the input and output tapers are modeled with two FDTD simulations, exciting an inner (c) and an outer (d) input port; by mirror symmetry, these two solves supply the transfer matrices for all four input and output tapers. The (e) common-mode rejection ratio (CMRR), (f) excess loss, and (g) phase error of the driven ports are computed from the hybrid cascade at the full device's own multimode section length of 159.3~\textmu m (167.3~\textmu m including both tapers). Because a multimode section this long is infeasible to simulate directly with FDTD, a reduced-length device (40~\textmu m multimode section) is instead simulated end-to-end and compared against the hybrid method at that same reduced length. Close agreement between the two methods is seen in both the (h) transmission and (i) relative phase of each output port, and (j) the difference between the hybrid and full FDTD relative phases remains small across the band, particularly for the output carrying the most power (output 3). The output ports are numbered from bottom to top and the MMI is excited from the top port.}
\label{fig:mmi_4x4}
\end{figure}

\begin{table}[tbp]
\centering
\caption{Geometric parameters of the 4$\times$4 MMI 90-degree hybrid (Fig.~\ref{fig:mmi_4x4}a).}
\label{tab:mmi_4x4_params}
\begin{tabular}{lll}
\toprule
Symbol & Description & Value \\
\midrule
$w_{\text{wg}}$ & Waveguide width & 0.35~\textmu m \\
$w_{\text{mm}}$ & Multimode section width & 8.5~\textmu m \\
$L_{\text{mm}}$ & Multimode section length (nominal) & 159.3~\textmu m \\
$w_{\text{taper}}$ & Taper width & 1.3~\textmu m \\
$L_{\text{taper}}$ & Taper length & 4.0~\textmu m \\
$d_{\text{wg}}$ & Waveguide separation & 2.125~\textmu m \\
\bottomrule
\end{tabular}
\end{table}

90-degree hybrids are vital components of integrated coherent receivers, enabling the extraction of in-phase and quadrature components of a modulated optical signal \citep{halir2011high}. A 4$\times$4 MMI coupler can be designed to implement this functionality \citep{yasui2021structural}, but the multimode section length required to establish the correct phase relationships across its four output ports typically far exceeds what can be reasonably simulated with FDTD alone. This device is therefore the clearest demonstration of the hybrid method's central advantage: an EME body of essentially any length costs the same handful of mode solves as a short one, while only the input and output tapers require FDTD. 

The 4$\times$4 90-degree hybrid's geometry is similar to the 2$\times$2 MMI splitter's, with four equally spaced input and output tapers in place of two (Fig.~\ref{fig:mmi_4x4}a); its parameters are listed in Table~\ref{tab:mmi_4x4_params}. Its solver architecture likewise mirrors the 2$\times$2 splitter's, with FDTD simulations covering the input and output tapers and an EME body covering the multimode section (Fig.~\ref{fig:mmi_4x4}b). A four-port device, however, is only mirror-symmetric between its inner pair of ports and its outer pair, not between every port, so a single FDTD simulation no longer determines the full transfer matrix. Two are required here: one exciting an inner input port (Fig.~\ref{fig:mmi_4x4}c) and one exciting an outer input port (Fig.~\ref{fig:mmi_4x4}d), from which the remaining two input tapers and all four output tapers follow by mirror symmetry.

The hybrid demonstrated here targets a 220~nm SOI platform with a 200~nm minimum feature size, as the 2$\times$2 splitters of Section~\ref{sec:2x2_mmi_splitter} do, making it feasible for almost any UV-lithography-based fabrication process. Unlike the other devices in this work, which operate over the C-band, this hybrid is designed and simulated over a section of the O-band, from 1.29 to 1.33~\textmu m, sampled at eleven equally spaced wavelengths. Its performance is evaluated against three metrics: the common-mode rejection ratio (CMRR), the excess loss, and the phase error of the driven ports (Fig.~\ref{fig:mmi_4x4}e,f,g), computed from the hybrid cascade at the full device's own multimode section length of 159.3~\textmu m (167.3~\textmu m including both tapers). The geometry was arrived at through a coordinate-search optimization: the multimode width, taper width, taper length, and multimode length were swept in turn, with the taper width re-matched to each candidate multimode width rather than held fixed, since a taper that does not widen with the body cannot be rescued by any multimode length. \texttt{solve\_length\_sweep()} made the multimode length itself free to re-optimize at every other point, evaluating all three metrics across the full length dependence from a single cached mode solve rather than a new simulation per candidate length. The resulting design falls short of both targets: the phase error reaches $|\Delta\phi| = 6.2^\circ$, exceeding the $|\Delta\phi|\leq5^\circ$ target quoted for this device class \citep{yasui2021structural,halir2011high} by roughly $1.2^\circ$, and the CMRR falls short of the accompanying $\text{CMRR}\geq20$~dBe target by roughly 1.5~dBe; further refinement of this device is out of the scope of this work.

The full hybrid's own multimode section length precludes a full-device FDTD simulation: an attempt at the full 159.3~\textmu m length was abandoned after three days of wall-clock time, with the fields still far from the decay threshold that terminates a converged run. A reduced-length 4$\times$4 MMI with a 40~\textmu m multimode section is short enough to simulate end-to-end with FDTD and compare directly against the hybrid method. We simulate this reduced-length device excited from the top-most input waveguide and compare the transmission and relative phase of each output port against the hybrid result at the same length (Fig.~\ref{fig:mmi_4x4}h,i). Both quantities agree closely between the two methods across the band, and the difference between the hybrid and full FDTD relative phases stays small at every wavelength, particularly for the output carrying the most power (output 3, Fig.~\ref{fig:mmi_4x4}j). It is this agreement, checked directly at a length short enough for FDTD to afford, that justifies trusting the hybrid cascade at the full device's own multimode length, where a full FDTD check is no longer an option.

This example demonstrates the hybrid approach's capacity to simulate large photonic devices by using FDTD only where it is needed, to model the critical taper regions, while letting EME carry the long multimode body at no added cost in solve count. Beyond the improved physical accuracy FDTD offers over EME at these transitions, FDTD's ability to represent an arbitrary, non-parameterized structure also opens the taper region itself to design: exciting a specific combination of body modes with a structure EME's mode-matched formalism cannot represent. This capability is what several inverse design techniques require in the first place, since the adjoint method backpropagates a field through the same volume it was forward-simulated in, and a taper with enough design freedom to be worth optimizing has too many degrees of freedom for EME's parameterized geometry to express \citep{lalaukeraly2013adjoint}. The hybrid cascade therefore opens a class of device the two methods could not reach separately: an inverse-designed taper region, simulated in FDTD, cascaded against an EME body long enough that simulating the whole device in FDTD would be intractable \citep{hiesener2026enhancing}.

\section{Conclusions}\label{sec:conclusions}

We present BEEF, a hybrid simulation framework that cascades independently-developed FDTD and EME solvers to simulate the electromagnetic response of integrated photonic devices. The framework implements this cascade over a common geometry interface, allowing any solver to occupy any position in the cascade, in any order. This lets a device be decomposed into sections simulated with whichever solver suits each region best: FDTD where the geometry is complex or the response is not well captured by the modes excited in the structure, and EME where the geometry is simple and the response is well described by the guided modes.

We validate this approach using a 220~nm SOI 2$\times$2 MMI splitter, a directional coupler, an HCG Fabry--Perot resonator, and a 4$\times$4 MMI-based 90-degree optical hybrid. The standard 2$\times$2 MMI splitter agrees with full-device FDTD to within 0.59\% (bar) and 0.76\% (cross) across the C-band. A tapered variant demonstrates an EME body composed of multiple cross-sections and agrees with full-device FDTD to within 0.87\% (bar) and 2.01\% (cross). This larger discrepancy is consistent with the additional taper-transition reflections introduced by the longer chain of sections. A standalone EME simulation, which must additionally represent the input and output tapers within a truncated guided-mode basis, deviates from full-device FDTD by 3.69\% (bar) and 2.49\% (cross). The directional coupler demonstrates the effectiveness of calibrating the EME propagation constants against FDTD: the cut-position drift in the extracted coupling length decreases 167-fold. The \texttt{Circuit} class then combines this coupler with two HCG reflectors and two connecting waveguides through the Redheffer star product. The resulting model reproduces the measured Fabry--Perot resonance lineshape and free spectral range while allowing the entire ($L_{\text{DC}}$, $L_{\text{arm}}$) design plane to be explored through closed-form post-processing of four cached simulations, without additional FDTD or mode solves.

The 4$\times$4 MMI-based 90-degree optical hybrid directly demonstrates the central motivation of this framework. With a multimode-section length of 159.3~\textmu m, this device is computationally impractical to simulate using full-device FDTD. BEEF instead confines FDTD to the critical input and output tapers while using EME to propagate through the long multimode body. A reduced-length variant with a 40~\textmu m multimode section, short enough to simulate end-to-end with FDTD, confirms close agreement with the hybrid result in both transmission and relative output phase. This decomposition also provides a path toward devices that combine complex or inverse-designed FDTD taper regions with EME bodies that can be swept rapidly, enabling efficient exploration of large design spaces and new operating regimes.

The computational cost of the three approaches is driven by different quantities, so no single runtime ratio characterizes the difference between them. FDTD cost is set by the domain volume and the number of time steps, with the Courant condition tying the time step to the smallest feature that must be resolved \citep{taflove2005computational}, though a single broadband run returns the entire spectrum. EME cost is set by the number of distinct cross-sections, the modes retained at each interface, and the number of wavelength points \citep{gallagher2003eigenmode}. The hybrid inherits both behaviors, confining FDTD to the short end sections whose cost is fixed by the section size rather than the device length, while paying EME's per-wavelength mode solve for the body. It also carries overheads of its own: the $n_{\text{eff}}$ calibration adds a short FDTD simulation per body cross-section, and a section whose ports or modes are not related by symmetry requires a separate FDTD excitation for each. However, mode solves and section transfer matrices can be cached and reused across lengths, cut positions, and design-plane points, making the length and design-plane sweeps closed-form post-processing rather than resimulation. An FDTD result admits no equivalent reuse, since it is tied to the whole domain and any change in length changes that domain. The advantage is therefore largest for the devices demonstrated here, whose bodies are long and uniform. A like-for-like runtime comparison would additionally have to fix a common accuracy target, since the two methods converge in different parameters (grid resolution for FDTD, mode and cell count for EME).

Development of BEEF is continuing in expanding the example library, adding solver backends, and extending the core framework. The maintained repository includes every hybrid and EME example presented in this work, together with further devices not shown here including a 1$\times$2 MMI splitter, a 1$\times$4 MMI splitter, and a ring resonator and will be updated as additional devices are developed. The results presented here validate the common geometry and solver interfaces using Meep for FDTD and MEOW for EME. This provides a basis for adding backends such as Lumerical and FDTDX, including GPU-accelerated implementations that would reduce the cost of the FDTD sections directly, and for extending the existing Tidy3D support beyond mode solving. These engine-level extensions are complemented by new design strategies, including the application of inverse design to the FDTD regions of a hybrid cascade, where restricting the optimization domain to a short input makes the hundreds to thousands of required solves affordable. This sectioning premise is not specific to photonics: microwave and antenna analysis has long decomposed a structure into regions, solved each region with a full-wave method, and recombined them through generalized scattering matrices \citep{eleftheriades1994some,shi2025generalized}, so the same solver-agnostic interface carries over to RF devices and systems. By deriving each section's response from a first-principles electromagnetic solution rather than a fitted compact model, BEEF offers a practical route toward simulating photonic devices whose length, geometric complexity, or design freedom put them beyond the reach of any single existing method.

\section*{Declaration of competing interest}

The authors declare no conflicts of interest.

\section*{Code availability}

The BEEF source code is available at \url{https://github.com/jakeh13/BEEF}.

\section*{Data availability}

Data underlying the results presented in this paper are not publicly available at this time but may be obtained from the authors upon reasonable request.

\section*{Acknowledgements}

This material is based upon work supported in part by the National Science Foundation (NSF) Center ``EPICA'' under Grant No.~2052808, \url{https://epica.research.gatech.edu/}. Any opinions, findings, and conclusions or recommendations expressed in this material are those of the author(s) and do not necessarily reflect the views of the NSF. This research was supported in part through research cyberinfrastructure resources and services provided by the Partnership for an Advanced Computing Environment (PACE) at the Georgia Institute of Technology. The authors thank GlobalFoundries for providing silicon fabrication through the Fotonix\textsuperscript{TM} university program. The authors used a generative AI tool (Claude, Anthropic) to assist with editing and proofreading the text of this manuscript. The authors reviewed and edited all AI-assisted content and take full responsibility for the content of this publication.

\bibliographystyle{plainnat}

\bibliography{references}

\end{document}